\documentclass{iopart}
\usepackage{graphicx}
\usepackage{color}
\begin{document}

\title[Choosing a vacuum state in a spherical spacetime with a CKV]{Choosing a vacuum state in a spherical spacetime with a  conformal Killing vector}

\author{A.M.Venditti$^1$, C.C.Dyer$^2$}

\address{Department of Physics, University of Toronto, 50 St. George Street, Toronto, Ontario M5S 3H4, Canada$^1$}
\address{Department of Astronomy and Astrophysics, University of Toronto, 50 St. George Street, Toronto, Ontario M5S 3H4, Canada$^2$}
\address{Department of Physical and Environment Sciences, University of Toronto at Scarborough, 1265 Military Trail, Scarborough, Ontario M1C 1A4, Canada$^2$}
\eads{\mailto{avenditt@physics.utoronto.ca$^1$}, \mailto{dyer@astro.utoronto.ca$^2$}}

\begin{abstract}
We consider the problem of picking a physically motivated vacuum state on a spherically symmetric spacetime with an extra conformal Killing vector, as opposed to an extra Killing vector as in the Schwarzschild case.  Considering a conformal symmetry instead of a symmetry allows us to consider spacetimes that are dynamical and not static (like Schwarzschild).  The extra conformal symmetry allows us to calculate the response of particle detectors.  We present vacuum state that follows mathematically from the assumptions of regularity of the state at $r = 0$, conformal symmetry under translations along the conformal Killing vector, the equal time canonical commutation relations, and analyticity in the complex plane.  We look at the specific example of a self-similar LTB spacetime that represents a spherically symmetric but inhomogeneous cosmology.  We remark that the above procedure might be applied to a spherically symmetric collapse solution that represents black hole formation so that one can calculate the detailed spectrum of Hawking radiation during a collapse.
\end{abstract}

\pacs{04.62.+v, 04.70.Dy}

\section{Introduction}

In order to calculate the response of particle detectors following a given trajectory in a dynamic spacetime, one often picks spacetimes that have a large amount of symmetry.  Symmetries of the spacetime make PDE's such as the Klein-Gordon equation separable, allowing one to give a complete set of solutions to the PDE in terms of solutions to several ODE's.  The correlation function can be obtained from the complete set of solutions and hence the response of an Unruh-Dewitt detector \cite{Unruh_1976} can be calculated.\\

The response of detectors in anisotropic Bianchi spacetimes can be calculated due to the existence of three translational Killing vector fields \cite{Birrell_Davies_1982}.  In the eternal Schwarschild spacetime, existence of spherical symmetry and a global time translation Killing vector field allow these calculations to be done everywhere \cite{Birrell_Davies_1982}, \cite{Parker_Toms_2009}.\\

In \cite{Hawking_1975} black holes without the global time translation Killing vector field were considered.  It was demonstrated that all black holes formed from collapsing matter radiate with the characteristic black body spectrum.  However, the trade off for making an argument for black hole solutions formed from a general spherically symmetric matter is that the spectrum of radiation can only be calculated on future null infinity.\\

The problem with requiring more symmetry is that it often makes the physical system less realistic.  For example, the global time translation Killing vector in the Schwarzschild spacetime specifies a black hole that was not formed at a finite time in the past.  Hence, by assuming this symmetry one would not be able to obtain a physically realistic response of particle detectors away from future null infinity for black holes formed from collapsing matter.  In the cosmological realm, homogeneity is assumed to be true only at the largest scales and it is likely that at smaller scales the geometry of the universe is more like that of a Swiss Cheese model \cite{Dyer_1979}.  The swiss cheese model contains inhomogeneous regions such as a Lemaitre-Tolman-Bondi (LTB) spacetime.\\ 

We give a physically well motivated prescription for choosing a vacuum state that can be applied to any conformally coupled quantum field in a general spherically symmetric spacetime with an extra conformal Killing vector field.  This choice of vacuum state follows mathematically from the assumptions of regularity of the state at $r = 0$, conformal symmetry under translations along the conformal Killing vector, the equal time canonical commutation relations, and analyticity in the complex plane.\\ 

The reason for considering this type of spacetime is that it is more physical than considering a spherically symmetric solution with a time-like Killing vector (Schwarzschild) which is completely non-dynamical.  The late time response of comoving detectors is given for the self-similar LTB spacetime given in \cite{Dyer_1979}.\\

\section{A self-similar LTB spacetime}

This section is an overview of the results given in \cite{Dyer_1979} that are needed for this paper (for a full discussion one should consult \cite{Dyer_1979}). \\

A model for an inhomogeneous, spherically symmetric universe filled with pressure-less dust is given by the following LTB metric.
\begin{equation}
\label{eq:ch5_comovemetric}
ds^2 = dt^2 - \exp(2\lambda)dr^2 - r^2S^2d\Omega^2
\end{equation}
where $\lambda = \lambda(t,r)$ and $S = S(t,r)$.  The stress energy tensor is
\begin{equation}
T_{ab} = \rho u_{a}u_{b}
\end{equation}
where $\rho$ is the density and $u_{a}$ is the 4-velocity of the dust.\\

To solve the Einstein equations a self-similarity is imposed by making $\lambda = \lambda(t/r)$ and $S = S(t/r)$.  The self-similarity implies the existence of a vector field ($\xi_{a}$) satisfying the equation
\begin{equation}
\xi_{a || b} + \xi_{b || a} = 2g_{ab}
\end{equation}
where $||b$ denotes covariant differentiation.  $\xi_{a}$ is called a homothetic Killing vector field.  Self-similarity and the Einstein field equations imply that $S$ and $\lambda$ satisfy the following equations
\begin{eqnarray}
\label{eq:ch5_Sequation} S'^2 &=& 2E + 2/S \\
\label{eq:ch5_lambdaequation} \exp(\lambda) &=& (S - sS')/\sqrt{1 + 2E}
\end{eqnarray}
where $s = t/r$ and $S' = \frac{dS}{ds}$. E is an integration constant representing the energy of the shell in the infinite past.  The case $E=0$ will be used.  The solution to equation (\ref{eq:ch5_Sequation}) in the $E=0$ case is
\begin{equation}
\label{eq:ch5_EzeroS} s + g = \pm\frac{\sqrt{2}}{3}S^{3/2}
\end{equation}
where $g$ is an integration constant representing the degree of inhomogeneity in the metric.\\

In the $t$, $r$ coordinates the line element appears to be singular at $r=0$ \cite{Dyer_1979}.  Since the behaviour of the Klein-Gordon field will be needed near $r=0$ a coordinate transformation will be made to the coordinate $\omega$ given by $r = 2\omega^3/9t_{0}^2$, where $t_{0}$ is some arbitrary time scale.  In the $\omega$ coordinate the line element is
\begin{equation}
\label{eq:ch5_ltbmetric}
ds^2 = dt^2 - \left(\frac{t + \tilde{\gamma} \omega^3}{t_{0}}\right)^{4/3}\left[ \left( \frac{t + 3\tilde{\gamma}\omega^3}{t + \tilde{\gamma}\omega^3}\right)^2d\omega^2 + \omega^2 d\Omega^2 \right]
\end{equation}
where $\tilde{\gamma} = 6g/(27t_{0}^2)$.\\

The coordinates take on the values $\theta \in (0, \pi)$, $\varphi \in (0, 2\pi)$, $\omega \in (0, \infty)$ and $t \in (-\infty, \infty)$.  There is a space-like singularity at the set of points $t/\omega^3 = -\tilde{\gamma}$ which can be thought of as an inhomogeneous big bang singularity.  We will later introduce the coordinate $x$, defined by $x^3 = \omega^3/(t_0^2 t)$ which can take on values in $(-\infty, \infty)$, however for our purposes it is only necessary to discuss observers such that $t > 0$ so we can restrict $x$ to $(0, \infty)$. \\

It is generally true that for a spacetime with a conformal vector field (CKV) there exists a conformally transformed spacetime in which that C.K.V. becomes a Killing vector field.  To prove this consider a spacetime metric $g_{ab}$ with a conformal Killing vector field satisfying
\begin{equation}
L_{k}g_{ab} = \phi(x^c)g_{ab}
\end{equation}
where $L_{k}$ is the Lie derivative along the trajectory of the vector field $k^{a}$.  Consider a parameter $\tau$ along the trajectory of the vector field $k^{a}$ so that $k^{a}$ can be represented by the derivative $\frac{d}{d\tau}$.  The conformally transformed spacetime given by 
\begin{equation}
\bar{g}_{ab} = \exp\left(-\int{d\tau \phi(\tau)}\right)g_{ab}
\end{equation}
is a solution of the equation
\begin{equation}
L_{k}\bar{g}_{ab} = 0
\end{equation}
confirming that $k^{a}$ is a Killing vector field in the spacetime with metric $\bar{g}_{ab}$.\\

A coordinate can be chosen along the trajectories of the Killing vector field $k^{a}$ (say $\tau$) so that the metric is independent of this coordinate.  In a spherically symmetric, 4 dimensional spacetime this guarantees that the Klein-Gordon equation is separable and hence one can solve for a complete set of solutions.\\ 

The spacetime described above has a conformal Killing vector field given by
\begin{equation}
\label{eq:ch5_kv} \xi^{a}\frac{\partial}{\partial x^{a}} = t\frac{\partial}{\partial t} + \frac{\omega}{3}\frac{\partial}{\partial \omega}
\end{equation}
The coordinate $y$ will be chosen so that $\frac{d}{dy}$ is the conformal Killing vector field.  The relationship to the $t$ and $\omega$ coordinates can be obtained by equating (\ref{eq:ch5_kv}) and the relation
\begin{equation}
\frac{d}{dy} = \frac{\partial \omega}{\partial y}\frac{\partial}{\partial \omega} + \frac{\partial t}{\partial y}\frac{\partial}{\partial t}
\end{equation}
We obtain 
\begin{equation}
\omega = A\exp(y/3) \;\;\;\;\;\; t = B\exp(y)
\end{equation}
where $A$ and $B$ are functions of $x$ that are independent of $y$.  We choose $A = t_{0}x$ and $B = t_{0}$, where x will be another new coordinate.  The coordinate $x$ is constant along curves generated by the self-similarity transformation.  \\

The line element in the $y$, $x$ coordinates is given by
\begin{equation}
\label{eq:ch5_metric} ds^2 = t_{0}^2\exp(2y)\left[ dy^2 - \frac{b^2}{a^{2/3}}(dx + \frac{x}{3}dy)^2 - x^2a^{4/3}d\Omega^2\right]
\end{equation}
where $a \equiv 1+\gamma x^3$ and $b \equiv 1+3\gamma x^3$ with $\gamma = 6g/27$.  The $\exp(2y)$ factor is the only dependence of the metric on $y$.  Therefore, $d/dy$ is a conformal Killing vector in the above metric $g_{ab}$ and a Killing vector in the metric $\bar{g}_{ab}$ given by
\begin{equation}  
\label{eq:ch5_confequation} \bar{g}_{ab} = \exp(-2y)g_{ab}
\end{equation}
The conformal Killing vector $d/dy$ becomes null ($g(d/dy, d/dy) = 0$) when the coefficient of the $dy^2$ term in (\ref{eq:ch5_metric}) vanishes.  The surface where the C.K.V. $d/dy$ becomes null defines a conformal Killing horizon \cite{Dyer_Honig_1979}.  So there is a conformal Killing horizon at the solutions of the following equation
\begin{equation}
\label{eq:ch5_ckh}
1- \frac{b^2}{a^{2/3}}\frac{x^2}{9} = 0
\end{equation}
In order to count the number of positive roots in $x$ of the above equation we first factor the left hand side using the difference of squares.
\begin{equation}
\left( 1 - \frac{b}{a^{1/3}}\frac{x}{3}\right)\left(1 + \frac{b}{a^{1/3}}\frac{x}{3}\right) = 0
\end{equation}
The factor on the right is always positive for $\gamma > 0$ and
therefore has no roots for $x>0$.  Making the substitution $z = x^3$ in
the factor on the left and simplifying gives that the positive roots of
(\ref{eq:ch5_ckh}) are the same as the positive roots of the following equation.
\begin{equation}
\label{eq:ch5_polyckh}
27\gamma^3z^4 + 27\gamma^2z^3 + 9\gamma z^2 + (1 - 27\gamma) z - 27 = 0
\end{equation}
There is only one sign change in the coefficients of the polynomial on
the left hand side of (\ref{eq:ch5_polyckh}).  Therefore there can be at
most one positive real root.  Using $z = x^3$ we see that there is only one positive real root for $x>0$ and hence only one conformal Killing horizon in the expanding, inhomogeneous universe ($t>0$).\\

We now introduce the mathematical tools to study the evolution of a bundle of null geodesic curves in the above spacetime.  The purpose is to find other types of horizons.  The scalar expansion of a congruence of geodesic curves is defined to be
\begin{equation}
\theta = u^{a}_{||a}
\end{equation}
where $u^a$ is a tangent vector to the congruence of curves given by $d/d\lambda$ where $\lambda$ is an affine parameter along the curves and $||a$ is the covariant derivative with respect to the coordinate $x^a$ \cite{Wald_1994}. This quantity defines how the cross-sectional area of a small bundle of curves changes.  For example, a bundle of curves representing radially outgoing light rays in Minkowski spacetime would have positive expansion because the curves move away from each other as they move to increasing radius, while the expansion for radially ingoing light rays in Minkowski spacetime would be negative.\\

A trapped surface of a spherically symmetric spacetime is defined as the surface where the expansion is negative for both ingoing and outgoing sets of null geodesics \cite{Poisson_2004}.  The apparent horizon is the boundary of all the trapped surfaces, meaning that the expansion is zero for one set of radial null geodesics and negative for the other.  For example, the event horizon of the Schwarzschild spacetime is an apparent horizon \cite{Poisson_2004}.\\

It can be shown that the conformal Killing horizon (CKH) is also an apparent horizon of the spacetime (\ref{eq:ch5_ltbmetric}) by calculating the expansion of the congruence of the two independent, future-directed, radially moving, null geodesics.  The ingoing and outgoing tangent vectors, denoted by $k^a_-$ and $k^a_+$ respectively, are calculated by solving the equations
\begin{eqnarray}
\label{eq:ch5_nullgeodesic}k^bk^a_{||b} &=& 0 \\
\label{eq:ch5_nullnorm} k^ak_a &=& 0
\end{eqnarray}  
For convenience we will work in the conformally transformed spacetime $\bar{g}_{ab}$ given by (\ref{eq:ch5_confequation}).  We can work in the conformally transformed spacetime because null geodesics remain null geodesics under conformal transformations.  The tangent vectors in affine parametrization in the original spacetime can be obtained by a conformal transformation even though working in the barred spacetime ($\bar{g}_{ab}$) changes the geodesics from affine parametrization to non-affine parametrization. Taking the null geodesics to be defined by an affine parameter $\lambda$ in $\bar{g}_{ab}$ and working in the coordinates of (\ref{eq:ch5_ltbmetric}) we obtain the following expression for (\ref{eq:ch5_nullnorm})
\begin{equation}
\left(\dot{y} - \frac{b}{a^{1/3}}(\dot{x} + \frac{x}{3}\dot{y})\right)\left(\dot{y} + \frac{b}{a^{1/3}}(\dot{x} + \frac{x}{3}\dot{y})\right) = 0
\end{equation}
with
\begin{equation}
k^a = (\dot{y}, \dot{x}, 0, 0)
\end{equation}
where $\dot{y} \equiv dy/d\lambda$.  Setting each of the factors to zero gives us an equation representing one of the two independent null geodesics.
\begin{equation}
\label{eq:ch5_nullnormred}
\dot{y} \pm \frac{b}{a^{1/3}}(\dot{x} + \frac{x}{3}\dot{y}) = 0
\end{equation}
The geodesic equations for the null vector (\ref{eq:ch5_nullgeodesic}) reduce to the following expression
\begin{equation}
\label{eq:ch5_nullgeodesicred}
\dot{y} - \frac{b^2x}{3a^{2/3}}(\dot{x} + \frac{x}{3}\dot{y}) = C
\end{equation}
where $C$ is an integration constant.  Solving equations (\ref{eq:ch5_nullnormred}) and (\ref{eq:ch5_nullgeodesicred}) we obtain an expression for $k^a_{\pm}$ in the conformally transformed spacetime
\begin{equation}
k^a_{\pm} = C\left((1+xb/3a^{1/3})^{-1}, \mp \frac{a^{1/3}}{b}, 0, 0\right)
\end{equation}
It can be verified that in the original spacetime the tangent vectors given by the derivative with respect to an affine parameter are given by
\begin{equation}
k^a_{\pm} = Ce^{-2y}\left((1+xb/3a^{1/3})^{-1}, \mp \frac{a^{1/3}}{b}, 0, 0\right)
\end{equation}

It must also be determined where in the spacetime that the null vectors $k^a$ are future pointing: the reason is because if we find that the expansion associated with $k^a$ is negative for example but it is past pointing then as time runs forward the expansion is actually positive. To accomplish this the $k^a$ vectors are expressed in the ($t$, $\omega$) coordinates of (\ref{eq:ch5_ltbmetric}).  This is done because both the vectors $d/dy$ and $d/dx$ coordinates are time-like in certain regions while $d/d\omega$ is space-like everywhere and $d/dt$ is time-like everywhere.  The $d/dt$ component of the null vectors are
\begin{equation}
k^{0}_{\pm} = Ct\frac{e^{2y}}{1\pm xb/3a^{1/3}}
\end{equation}
So while the component $k^0_{+}$ is always positive for $x>0$, the component $k^{0}_{-}$ becomes negative when $1-xb/3a^{1/3}<0$.  Hence the vector $k^a_{+}$ is always future pointing but $k^{a}_{-}$ becomes past pointing when $1-xb/3a^{1/3} = 0$.  Note that this is also the value of $x$ where the CKH is found.\\

The expansion of each set of null geodesics is found to be given by the expressions
\begin{equation}
\theta_{\pm} = \frac{2e^{2y}(2x\mp3a^{1/3})}{(3a^{1/3}+xb)xa^{2/3}}
\end{equation}
It can be verified that for $\gamma > 8/27$ and $x>0$, $\theta_+ < 0$ and $\theta_- > 0$ everywhere.  However $k^a_-$ becomes past directed at the CKH so the true expansion associated with these null geodesic congruences becomes negative.  Hence, the CKH is also an apparent horizon for $\gamma > 8/27$.  This will be of importance later when we use the existence of the CKH to give a low frequency cut-off for the two point function.

\section{Conformally coupled Klein-Gordon field}
The action for a conformally coupled Klein-Gordon (KG) field in 4 dimensional spacetime is
\begin{equation}
\label{eq:ch5_kgaction}
S = \int{d^4x\sqrt{-g}\left(g^{ab}\partial_a \psi \partial_b \psi - \frac{1}{6}R\psi^2\right)}
\end{equation}
from which the equations of motion are derived as
\begin{equation}
\label{eq:ch5_kgequation} \frac{1}{\sqrt{-g}}\partial_{a}\left(\sqrt{-g}g^{ab}\partial_{b}\psi\right) + \frac{1}{6}R\psi = 0
\end{equation}
where $R$ is the Ricci scalar given by
\begin{equation}
R = -\frac{4}{3}\frac{27\gamma^2x^6 + 18\gamma x^3 + 1}{ab}
\end{equation}

The conformally coupled KG field has the property that if $\psi$ satisfies (\ref{eq:ch5_kgequation}) in the metric $g_{ab}$ then the field $\bar{\psi} = \phi^{-1}\psi$ will satisfy the KG equation in the conformally transformed metric $\bar{g}_{ab} = \phi^2g_{ab}$.  We can solve for the conformal fields in the metric $\bar{g}_{ab}$ given by (\ref{eq:ch5_confequation}) because the KG equation is separable there and then obtain the fields in the metric $g_{ab}$ by a conformal transformation.\\

In the metric where $d/dy$ is a Killing vector, the following ansatz is used.
\begin{equation}
\label{eq:ch5_modes} \bar{\psi}_{\nu l m} = \exp(- i \nu y)f_{\nu l}(x)Y_{lm}(\theta, \varphi)
\end{equation}
where $Y_{lm}(\theta, \varphi)$ are the spherical harmonics.  The vacuum of the ansatz (\ref{eq:ch5_modes}) is invariant under translations along the Killing vector $d/dy$ (because the modes are multiplied by a phase under shifts in $y$).  Substituting (\ref{eq:ch5_modes}) in the KG equation the following differential equation for $f_{\nu l}$ follows.

\begin{eqnarray}
\label{eq:ch5_fode}
\fl  &\left(\frac{x^2}{9} - \frac{a^{2/3}}{b^2}\right)abx^2\frac{d^2 f_{\nu l}}{dx^2} 
  + \left(\frac{2i\nu}{3}abx^3 + \frac{d}{dx}\left(\left(\frac{x^2}{9}-\frac{a^{2/3}}{b^2}\right)abx^2\right)\right)\frac{df_{\nu l}}{dx} \\
\fl \nonumber &+ \left(-\nu^2abx^2 + \frac{i \nu}{3}\frac{d}{dx}(abx^3) + l(l+1)\frac{b}{a^{1/3}} - \frac{2}{9}x^2(27\gamma^2x^6 + 18\gamma x^3 + 1)\right)f_{\nu l} = 0
\end{eqnarray}

The singularities at $x=0$ are regular; the powers of $x$ appearing the equation \ref{eq:ch5_fode} have a finite negative power.  Therefore (\ref{eq:ch5_fode}) can be solved perturbatively by expressing the coefficients as a series expanded around $x=0$.  The two independent solutions take the form

\begin{eqnarray}
\label{eq:ch5_regsol} f_{\nu l} &=& x^l\sum_{0}^{\infty}a_n x^n \\
\label{eq:ch5_singsol} f_{\nu l} &=& x^{-l-1}\sum_{0}^{\infty}b_n x^n
\end{eqnarray}  
where $l>0$.  The solution (\ref{eq:ch5_regsol}) is regular at $x=0$ while (\ref{eq:ch5_singsol}) is singular there.  The line element (\ref{eq:ch5_metric}) at $x=0$ reveals that the metric is regular there.  The $d\Omega^2$ term in (\ref{eq:ch5_metric}) is multiplied by $x^2$, so it does not contribute at $x=0$.  The same behaviour occurs in flat spacetime in spherical coordinates at $r=0$ even though it is a regular point, therefore we expect the spacetime to be completely well behaved at $x=0$.  Since the metric is well behaved at $x=0$, it is expected that the modes will be also.  Hence, only equation (\ref{eq:ch5_regsol}) will be used for the $f_{\nu l}$ part of the modes.\\

Note that the coefficients to the derivatives of $f_{\nu l}$ in equation (\ref{eq:ch5_fode}) all have a finitely sized radius of convergence given by the radius of convergence of the fractional powers $a^{2/3}$ and $a^{1/3}$.  The value of any function given in a local neighbourhood around $x = 0$ can be given uniquely by analytic continuation at any other $x$ greater than the conformal Killing horizon or the radius of convergence of the fractional powers above.  The fact that the solution does not have a well-defined value at the conformal Killing horizon is similar to the Schwarzschild case where both the Unruh vacuum \cite{Unruh_1976} and the Boulware vacuum \cite{Boulware_1975} are similarly undefined at the future horizon.\\

Requiring a vacuum that is invariant along the Killing trajectory reduced the arbitrariness down to two complex constants ($b_{0}$ and $a_{0}$ in (\ref{eq:ch5_regsol}) and (\ref{eq:ch5_singsol})).  Imposing regularity at $x=0$ implied $b_{0} = 0$, reducing the choice of vacuum down to one complex constant ($a_{0}$).  Finally, the canonical commutation relations of the KG field operator imply the modes must satisfy the following normalization.
\begin{eqnarray}
\fl \label{eq:ch5_normalization} (\psi_{\nu' l' m'}, \psi_{\nu l m}) &=& \int{d^3x\sqrt{-g}g^{0a}\left(\psi_{\nu' l' m'}^*\partial_{a} \psi_{\nu l m}-\psi_{\nu l m}\partial_{a}\psi_{\nu' l' m'}^*\right)} \\ 
 \nonumber &=&  i\delta(\nu - \nu') \delta_{l l'} \delta_{m m'}
\end{eqnarray}
where the integration is done over the $x^1 = y$, $x^2 = \theta$ and $x^3 = \varphi$ direction.  The normalization (\ref{eq:ch5_normalization}) fixes one of the two remaining real constants leaving only a phase factor unfixed.  The phase factor disappears when calculating correlation functions.  Hence imposing the conditions of regularity at $x=0$, invariance under $y$ translations and the normalization (\ref{eq:ch5_normalization}) uniquely specifies the vacuum state.\\

Note that the normalization (\ref{eq:ch5_normalization}) is done by integrating over the $y$, $\theta$ and $\varphi$ directions, i.e. it is done on a constant $x$ surface.  The normalization integral (\ref{eq:ch5_normalization}) is conserved and is independent of which $x$ surface we use.  The constant $x$ surfaces are space-like surfaces for $x$ greater than the conformal Killing horizon.  It does not matter that the $x$ surfaces are non-spacelike for certain values because due to conservation it is equivalent to normalizing on an $x$ surface that is space-like.  Also we wish to point out that using a constant $x$ surface for normalization means that the conjugate momentum for this system is defined as:
\begin{equation}
\Pi(x, y, \theta,\varphi) = \frac{\delta S }{\delta \left(\partial_{x}\psi(x, y, \theta, \varphi) \right)}
\end{equation}
where $S$ is the action (\ref{eq:ch5_kgaction}) and the canonical commutation relations are given by the Dirac quantization rule
\begin{equation}
[\hat{f}, \hat{g}] = i\widehat{\{f, g\}}
\end{equation}
which gives
\begin{equation}
\label{eq:ch5_ccrs}
[\hat{\psi}(x, y, \theta, \varphi), \hat{\Pi}(x, y', \theta', \varphi')] = i \delta(y - y')\delta(\theta - \theta')\delta(\varphi-\varphi')
\end{equation}
Note that the commutation relations are on a constant $x$ surface.  Also, the choice of $\Pi(x, y, \theta, \varphi)$ as a generalized momentum variable gives corresponding Hamilton's equations which of course are equivalent to the equations of motion for the $\psi$ field.  This gives us a Poisson bracket ($\{ , \}$) for use in the Dirac quantization rule. \\

To find the specific condition that the normalization condition (\ref{eq:ch5_normalization}) places on the field modes we note that if we choose a value for the $a_{0}$ term in the series (\ref{eq:ch5_regsol}) then we have uniquely specified the solution $f_{\omega l}$.  This follows from the fact that equation (\ref{eq:ch5_fode}) is linear and that the Frobenius method will specify each of the coefficients in the series in terms of $a_{0}$.  Hence we can write the coefficients as:
\begin{equation}
\label{eq:ch5_aequation}
a_{m} = c_{m}a_{0}
\end{equation}
where $m$ is a non-negative integer and the $c_{m}$ only depend on $\nu$, $l$ and $\gamma$ (the parameters appearing in (\ref{eq:ch5_fode})).  So we see from (\ref{eq:ch5_aequation}) that specifying $a_{0}$ will fix the rest of the $a_n$'s and hence fix the function $f_{\omega l}$.\\

For convenience we introduce the following function:
\begin{equation}
h_{\omega l} = \frac{f_{\omega l}}{a_0} = x^l \left(1 + \sum_{n = 1}^{\infty}c_n x^n\right)
\end{equation}
where we have used $c_0 = 1$ by definition.  Using the ansatz (\ref{eq:ch5_modes}) the normalization condition (\ref{eq:ch5_normalization}) reduces to the following condition on $a_{0}$.
\begin{equation}
\label{eq:ch5_a0condition}
\fl |a_{0}|^{2} = i\left(2\pi a b x^2 \left[ \frac{2i \omega x}{3}|h_{\omega l}|^2 + \left(\frac{x^2}{9} - \frac{a^{2/3}}{b^2}\right)\left(h^{*}_{\omega l}h'_{\omega l} - h_{\omega l}h^{* '}_{\omega l}\right)\right] \right)^{-1}
\end{equation}
The integration (\ref{eq:ch5_normalization}) is done over a constant $x$ surface so the condition (\ref{eq:ch5_a0condition}) is evaluated at some constant $x = u$. \\

From the expansion (\ref{eq:ch5_aequation}) we see that $f_{\omega l} =
a_{0}h_{\omega l}$ where $h_{\omega l}$ does not depend on $a_{0}$ and
hence does not depend on the normalization surface we choose.  Our
results should not depend on the normalization surface chosen hence it
follows that $\frac{df_{\omega l}}{du} = \frac{d a_{0}}{du} = 0$ where
$x = u$ is the normalization surface.  That $\frac{d a_{0}}{du} =
0$ follows from the fact that the inner product given in
(\ref{eq:ch5_normalization}) is conserved.  This is because the
normalization factor is simply $\sqrt{( h_{\omega l}, h_{\omega l})}$.
One can use equation \ref{eq:ch5_fode} to show that the right hand side of (\ref{eq:ch5_a0condition}) is indeed a constant; this derivation is given in the appendix.\\

Before we calculate the correlation function we note that the modes defining the vacuum state (given by $\psi_{\nu l m} = \exp(-y)\bar{\psi}_{\nu l m}$) are positive frequency with respect to $y$.  The $y$ coordinate is time-like inside the conformal Killing horizon and space-like outside of it.  This choice of coordinate is similar to \cite{Boulware_1975} where the author defines a vacuum state in the Schwarzschild space-time that is positive frequency with respect to the $t$ coordinate.  The $t$ coordinate is time-like outside of the horizon of the Schwarzschild space-time at $r = 2m$ and is space-like inside it.\\ 

The correlation function in the vacuum state defined above (the vacuum state which is regular at $r=0$ and has a conformal symmetry along $d/dy$) can be calculated by using the usual sum over all modes formula.  The general formula can be found in \cite{Birrell_Davies_1982}.
\begin{equation}
\fl G(x,x') = \sum_{l,m}\int_{0}^{\infty}{d\omega \exp(-i\omega(y-y')-(y+y'))f_{\omega l}(x)f^{*}_{\omega l}(x')Y_{l m}(\theta, \varphi)Y^{*}_{l m}(\theta', \varphi')}
\end{equation}
Only the correlation function evaluated along the points $\theta = \theta'$ and $\varphi = \varphi'$ will be of interest.  The following formula can be used to simplify the sum over modes:
\begin{equation}
\sum_{m = -l}^{m = l}Y_{l m}(\theta, \varphi)Y^{*}_{l m}(\theta, \varphi) = (2l+1)/4\pi
\end{equation}
giving
\begin{equation}
\label{eq:ch5_correlationfunc}
\fl G(x, x') = \sum_{l=0}^{\infty}\int_{0}^{\infty}{d\omega \exp(-i\omega(y-y')-(y+y'))f_{\omega l}(x)f^{*}_{\omega l}(x')\frac{2l+1}{4\pi}}
\end{equation}

Evaluating $|a_{0}|^2$ for at the lowest order in x, the following expression is obtained
\begin{equation}
|a_{0}|^2 = \frac{3}{4\pi(1+\gamma u^3)(1+3\gamma u^3)u^3\omega}
\end{equation}
where $u$ is the constant $x$ surface that the modes are normalized on.  The Taylor expansion for $f_{\omega l}$ is given by
\begin{equation}
f_{\omega l} = a_{0}x^l(1+O(x^2))
\end{equation}
It can be seen that there is a logarithmic divergence in the correlation function at $\omega = 0$ due to the normalization on the modes.  Logarithmic IR divergences occur in many other physical vacuum states, for example see the Bunch-Davies vacuum state on de Sitter spacetime \cite{Rajaraman_2010}.  To solve this a `brute-force' IR cutoff must be imposed.  This procedure is similar to one used to cure an IR divergence that occurs in de Sitter spacetime \cite{Xue_2011} where an IR cutoff is given by the location of the cosmological horizon.  The reason that the size of the horizon is used as an upper cutoff for wavelength is that there should not be particles whose wavelength extends  across a horizon into a region that is causally disconnected from the one which the particle is in.\\

A value for the cutoff is given by the constant $x$ surface that is the conformal Killing horizon.  This surface is also an apparent horizon, so similarly to the deSitter case in \cite{Xue_2011} it would be natural to impose that no wave mode have a wavelength that is larger than the apparent horizon.  The single solution to (\ref{eq:ch5_ckh}) is denoted by $x=\beta$.  The function $f_{\omega l}(x)$ has approximately the wavelength $2\pi/\omega$.  Therefore only modes with $\omega > 2\pi/\beta$ will be included in the sum to calculate the correlation function.
\begin{equation}
\fl G(x, x') = \sum_{l=0}^{\infty}\int_{2\pi/\beta}^{\infty}{d\omega \exp(-i\omega(y-y')-(y+y'))f_{\omega l}(x)f^{*}_{\omega l}(x')\frac{2l+1}{4\pi}}
\end{equation}
Performing the integration and summation gives
\begin{equation}
\fl G(x,x') = N(u, \gamma)\frac{Ei\left(1, \frac{2\pi}{\beta}(\epsilon +i(y-y')) \right)\left(\frac{xx'}{u^2}\right)^{1/4}P\left(\frac{1}{2},\frac{1}{2}, \frac{1 + xx'/u^2}{1-xx'/u^2}\right)}{\left(1-\frac{xx'}{u^2}\right)^{3/2}\exp(y+y')}
\end{equation}
where $N(u,\gamma)$ is only dependent on the normalization surface $x=u$ and the inhomogeneity parameter $\gamma$.  $Ei(1, x)$ is the exponential integral special function and $P(a, b, x)$ is the associated Legendre polynomial of the first kind.  They are given by the following.
\begin{eqnarray}
Ei(1, x) &=& \int_{1}^{\infty}{dy \exp(-y x) y^{-1}} \\
P(1/2, 1/2, x) &=& \frac{(x+1)^{b/2} \mathstrut_2 F_1(-a, a+1, 1-b, 1/2-1/2x)}{(x-1)^{b/2}\Gamma(1-b)}
\end{eqnarray}

The correlation function is being considered for small values of $x$ so it will be expanded around $x=0$ in a series.  Keeping only the leading term the following is obtained
\begin{equation}
\label{eq:ch5_centrecorrelation}
G(x,x') = N(u, \gamma)\frac{Ei\left(1, \frac{2\pi}{\beta}(\epsilon +i(y-y')) \right)}{\pi^{1/2}\exp(y+y')}
\end{equation}
The above function is the correlation function for an observer following the trajectory $x=0$.  It is also the leading order approximation to the correlation function for an observer at small $x$.\\

One might expect that the modes that define the vacuum state should oscillate infinitely rapidly as the apparent horizon is approached.  For example, the modes defining the Unruh vacuum state oscillate infinitely rapidly near the future horizon of the eternal Schwarzschild spacetime.  This is not the case for the other two standard vacuums on the Schwarzschild spacetime.  However the Unruh vacuum is thought to be the physically correct vacuum for an object collapsing to form a black hole.  In other words, it is the correct vacuum for a star that collapses to form a future horizon.\\

Solving (\ref{eq:ch5_fode}) numerically for regular initial conditions at $x=0$, it is found that the modes $f_{\omega l}(x)$ seem to oscillate infinitely rapidly at the conformal Killing horizon.  Hence the modes (\ref{eq:ch5_modes}) that specify the vacuum state have the physically desired property that they oscillate infinitely rapidly at the conformal Killing horizon.

\section{Response of Particle Detectors}
The response rate of an Unruh-Dewitt detector will be calculated for observers co-moving with the dust of the above LTB metric.  The expression for the response rate is given by
\begin{equation}
\label{eq:ch5_generalresponserate}
\dot{F}_{\tau}(\omega) = \Re\left(\int_{0}^{\tau - \tau_0}{ds \exp(-i \omega s) G( x(\tau) , x(\tau - s))}\right)
\end{equation}
$s$ is the amount of proper time passed since the initial time $\tau_{0}$ at which the detector is in its ground state.  $G(x, x')$ is the correlation function and $\tau$ is the time at which the response rate of the detector is evaluated.\\

The $r$ and $t$ coordinates of the metric (\ref{eq:ch5_comovemetric}) will be used to parametrize a co-moving observer.  $r$ is constant for a radially infalling dust particle and $t$ is the proper time of this dust particle.  The coordinate transformation $(x,y) \rightarrow (t, r)$ is given by
\begin{equation}
(x,y) = \left(\left(\frac{9r}{2t}\right)^{1/3}, \ln(t)\right)
\end{equation}
The trajectory of a comoving dust particle can therefore be parametrized by the $y$ coordinate.  Making the substitution $t = \exp(y)$ the integral becomes
\begin{equation}
\label{eq:ch5_responserate}
\fl \dot{F}_{y}(\omega) = \Re\left(\frac{1}{\exp(y)}\int_{y_{0}}^{y}{dy' \exp\left(-i \omega \left(e^y - e^{y'}\right)\right) Ei(1, \frac{2\pi}{\beta}(\epsilon + i(y-y')))}\right)
\end{equation}
where $y = \ln(\tau)$ is the time at which the detector response is evaluated and $y_0 = \ln(\tau_0)$ is the time at which the detector is taken to be in its ground state.  The correlation function for the observer at $x=0$ given by (\ref{eq:ch5_centrecorrelation}) was used.\\

It can be demonstrated that the response rate given by equation
(\ref{eq:ch5_responserate}) goes to zero as $y \rightarrow \infty$ for a
fixed initial time $y_0$.  This is true simply due to the fact that the
$\exp(-y)$ factor dominates over the integral part of the equation.  A
detailed proof of this result is given in the appendix.\\ 

We have therefore demonstrated that the response of a comoving particle
detector near $x = 0$ is a null response at late times in the vacuum we
have described previously.  The null response at late times confirms
that the vacuum state we have chosen is physically reasonable because
the spacetime (\ref{eq:ch5_ltbmetric}) approaches Minkowski spacetime at
late times.  The spacetime approaches Minkowski at late times in the
sense that the Ricci scalar $R$ or the Kretschmann scalar
$R_{abcd}R^{abcd}$ approach zero at late times, hence the curvature goes
to zero at late times.  Therefore any geodesic particle detector should
register zero particles at late times.\\

A similar argument can demonstrate that any comoving detector (constant $r, \theta, \phi$ trajectory) will not click.  This can be seen by using the relation $x = \left(9r/2t\right)^{1/3}$.  As $t \rightarrow \infty$, $x \rightarrow 0$.  The integral (\ref{eq:ch5_generalresponserate}) can be divided into two parts, one for late times or $ 0 < x < x_0$, and one for early times $x > x_0 > 0$.  The late time part of the response rate is given by (\ref{eq:ch5_responserate}) and hence becomes zero at late times as before.  The early time part of the response rate will also be exponentially damped by the $\exp(-y)$ factor and hence tend to zero as $y\rightarrow\infty$.  Therefore, the response rate for a comoving (constant $r$) particle detector will also tend to zero at late times.

\section{Discussion}
The selection of a vacuum state is a non-trivial process for a general background metric.  In the eternal Schwarzschild spacetime, there are three common choices for the vacuum state.  The Hawking-Hartle vacuum \cite{Hartle_Hawking_1976} is a state that is regular on the past and future horizons.  The Boulware \cite{Boulware_1975} vacuum is symmetric with respect to the time translation Killing vector field $\partial/\partial t$.  Finally the Unruh vacuum state \cite{Unruh_1976} is symmetric with respect to the Killing vector on the past horizon and symmetric with respect to the Killing vector on past null infinity.\\

We have given a physically motivated prescription for choosing the vacuum state of a conformally coupled quantum field on the spacetime given in \cite{Dyer_1979}.  More specifically, the vacuum state given in this paper follows mathematically from the assumptions of regularity of the state at $r = 0$, conformal symmetry under translations along the conformal Killing vector, the equal time canonical commutation relations, and analyticity in the complex plane.\\

The assumption of complex analyticity is used to continue the $x$ part of the wave modes around the conformal Killing horizon (see the discussion under equation (\ref{eq:ch5_regsol})).  The equal time commutation relations hold for $x$ greater than the conformal Killing horizon, which by the conservation of the commutation relations means that the equal $x$ commutation relations (\ref{eq:ch5_ccrs}) must hold.  Conformal symmetry along the CKV is a physically reasonable condition and allows us to separate the Klein-Gordon equation.  Finally, regularity of the vacuum state at $r=0$ is required for a physically reasonable matter field.\\

The above prescription for the vacuum state can be used for any conformally coupled field on any metric that is spherically symmetric, has an extra conformal Killing vector, and is regular at the centre of spherical coordinates.  The authors do not know of any current metrics satisfying these properties which could also represent a collapse to a black hole.  However, there are solutions to the Einstein Field Equations representing spherically symmetric matter collapsing to form a black hole \cite{Husain_1994}.  The solution in \cite{Husain_1994} also has a conformal Killing vector field.  Unfortunately the solutions given there have a time-like singularity at the centre of the spacetime which precludes the vacuum state prescription in this paper.\\

However, any solution of Einstein's equations representing matter collapsing to some central point ($r=0$) should be regular along the line of points specified by $r=0$ before some specified time when the singularity forms.  The metric is regular at these points; hence we expect the vacuum state to also be regular there which gives us one of the conditions we have used to specify the above vacuum state.  Further, we expect a conformally coupled field to have the same conformal symmetry as the background spacetime, which gives us the second condition on our vacuum state.  These two conditions, along with the commutation relations and complex analyticity, uniquely specify the vacuum state and for the above reasons we believe that it is the physically correct vacuum for spherically symmetric collapse solutions with a conformal Killing vector.\\

Finally, we note that the condition of regularity at $r=0$ cannot be used on the eternal Schwarzschild spacetime because there is always a singularity in the metric at $r=0$ and therefore there is no reason to expect that other scalars (such as a Klein-Gordon field) would not also blow up there.  By `eternal Schwarzschild spacetime' we mean the spherically symmetric spacetime that everywhere has a time translational Killing vector.  This spacetime cannot represent matter collapsing to form a black hole.  The conditions we have used to define a vacuum are more physically realistic in the sense that they apply to a spacetime representing a black hole that was formed at some finite time in the past.\\

\appendix
\section{Proof of null response of (\ref{eq:ch5_responserate})}

Here we demonstrate that the response rate given by (\ref{eq:ch5_responserate}) tends to zero as $y \rightarrow \infty $.  For convenience we make the following definition for the integrand of (\ref{eq:ch5_responserate}).
\begin{equation}
H(y, y') = \exp\left(-i \omega \left(e^y - e^{y'}\right)\right) Ei(1, \frac{2\pi}{\beta}(\epsilon + i(y-y')))
\end{equation}
So the response rate (\ref{eq:ch5_responserate}) becomes
\begin{equation}
\exp(-y)\Re\left(\int_{y_0}^{y}{dy'H(y,y')}\right)
\end{equation}
The following set of inequalities hold:
\begin{equation}
\fl 0 \leq \Re\left(\int_{y_0}^{y}{dy'H(y,y')}\right) \leq \left|\int_{y_0}^{y}{dy'H(y,y')}\right| \leq \int_{y_0}^{y}{dy'\left|H(y,y')\right|}
\end{equation}
where we have assumed that the response rate is non-negative.  The definition of the exponential integral function $Ei(1, z)$ for complex $z$ with $\Re(z) > 0$ is given by:
\begin{equation}
Ei(1, z) = \int_1^{\infty}{\frac{\exp(-az)}{a}da}
\end{equation}
Using this defintion we further have the following set of (in)equalities holding:
\begin{eqnarray}
\fl \int_{y_0}^{y}{dy'\left|H(y,y')\right|} &\leq& \int_{y_0}^{y}{dy'\int_1^{\infty}{da \left|\frac{\exp\left(-a\left(\frac{2\pi}{\beta}\left(\epsilon + i(y-y')\right)\right)\right)}{a}\right|}} \\
\fl \nonumber &=& \int_{y_0}^{y}{dy'\int_1^{\infty}{da \frac{\exp\left(-\frac{2\pi\epsilon a}{\beta}\right)}{a}}}\\
\fl \nonumber &=& (y-y_0)Ei(1,2\pi\epsilon/\beta)
\end{eqnarray}
Combining all these results we have that the response rate obeys the following inequality
\begin{equation}
 \fl 0 \leq \exp(-y)\Re\left(\int_{y_0}^{y}{dy'H(y,y')}\right) \leq \exp(-y)(y-y_0)Ei(1,2\pi\epsilon/\beta)
\end{equation}
which clearly demonstrates that the response rate goes to zero at late times ($y \rightarrow \infty$).\\

\section{Proof that the right hand side of \ref{eq:ch5_a0condition} is a constant}
Consider any function $h(x)$ which is a solution to (\ref{eq:ch5_fode}).  For convenience we define the following function:
\begin{equation}
\alpha(x) \equiv \left(\frac{x^2}{9} - \frac{a^{2/3}}{b^{2}}\right)
\end{equation}
In terms of $\alpha(x)$ equation (\ref{eq:ch5_fode}) is given by
\begin{eqnarray}
\label{eq:ch5_hode}
\fl \alpha a b x^2 \frac{d^2h}{dx^2} &+& \left(\frac{2i\nu}{3}abx^3 + \frac{d}{dx}\left(\alpha a b x^2\right)\right) \frac{dh}{dx} \\
 \fl \nonumber &+& \left(-\nu^2 ab x^2 + \frac{i \nu}{3}\frac{d}{dx}(abx^3) + l(l+1)\frac{b}{a^{1/3}} - \frac{2}{9}x^2(27\gamma^2 x^6 + 18\gamma x^3 + 1)\right) h = 0
\end{eqnarray}
Multiplying (\ref{eq:ch5_hode}) by $h^*$ (the complex conjugate of h) and then subtracting the complex conjugate of the resulting equation we end up with
\begin{eqnarray}
\label{eq:ch5_cchode}
\alpha abx^2\left(\frac{d^2h}{dx^2}h^* - h \frac{d^2h^*}{dx^2}\right) + \frac{d}{dx}(\alpha abx^2)\left(\frac{dh}{dx}h^* - h\frac{dh^*}{dx}\right) \\
 \nonumber  + \frac{2i\nu}{3}abx^3\left(\frac{dh}{dx}h^* + h\frac{dh^*}{dx}\right) + \frac{2i\nu}{3}\frac{d}{dx}(abx^3)|h|^2 = 0
\end{eqnarray}
The two terms of the first line of \ref{eq:ch5_cchode} are a total derivative and the two terms on the second line of \ref{eq:ch5_cchode} are also a total derivative.
\begin{equation}
\frac{d}{dx}\left(\alpha abx^2\left(\frac{dh}{dx}h^* - h\frac{dh^*}{dx}\right)\right) + \frac{2i\nu}{3}\frac{d}{dx}\left(abx^3|h|^2\right) = 0
\end{equation}
After factoring we finally end up with:
\begin{equation}
\frac{d}{dx}\left(abx^2\left(\frac{2i\nu x}{3}|h|^2 + \alpha\left(h^*\frac{dh}{dx} - h\frac{dh^*}{dx}\right)\right)\right) = 0
\end{equation}
which means
\begin{equation}
abx^2\left(\frac{2i\nu x}{3}|h|^2 + \alpha\left(h^*\frac{dh}{dx} - h\frac{dh^*}{dx}\right)\right)
\end{equation}
is a constant.  This means that the right hand side of equation (\ref{eq:ch5_a0condition}) is a constant.\\

\bibliographystyle{unsrt}
\bibliography{myrefs}
\end{document}